\documentclass{article}
\usepackage{graphicx}
\usepackage{multirow}

\title{Statistical Downscaling of Temperature Distributions from the Synoptic Scale to the Mesoscale Using Deep Convolutional Neural Networks}

\author{
Tsuyoshi Thomas SEKIYAMA\\
Meteorological Research Institute\\
1-1 Nagamine, Tsukuba, Ibaraki 3050052 Japan\\
\texttt{tsekiyam@mri-jma.go.jp}\\
}

\begin{document}
\maketitle

\begin{abstract}
Deep learning, particularly convolutional neural networks for image recognition, has been recently used in meteorology. One of the promising applications is developing a statistical surrogate model that converts the output images of low-resolution dynamic models to high-resolution images. Our study exhibits a preliminary experiment that evaluates the performance of a model that downscales synoptic temperature fields to mesoscale temperature fields every 6 hours. The deep learning model was trained with operational 22-km gridded global analysis surface winds and temperatures as the input, operational 5-km gridded regional analysis surface temperatures as the desired output, and a target domain covering central Japan. The results confirm that our deep convolutional neural network (DCNN) is capable of estimating the locations of coastlines and mountain ridges in great detail, which are not retained in the inputs, and providing high-resolution surface temperature distributions. For instance, while the average root-mean-square error (RMSE) is 2.7 K between the global and regional analyses at altitudes greater than 1000 m, the RMSE is reduced to 1.0 K, and the correlation coefficient is improved from 0.6 to 0.9 by the surrogate model. Although this study evaluates a surrogate model only for surface temperature, it probably can be improved by augmenting the downscaling variables and vertical profiles. Surrogate models of DCNNs require only a small amount of computational power once their training is finished. Therefore, if the surrogate models are implemented at short time intervals, they will provide high-resolution weather forecast guidance or environment emergency alerts at low cost.
\end{abstract}

\section{Introduction}
High-resolution dynamic models require much greater use of computational resources than low-resolution models. Therefore, downscaling procedures have been in demand for meteorological or climatological simulations to infer high-resolution variables from low-resolution fields. Statistical downscaling is preferable to dynamic downscaling, especially when computational time or resources are limited, because statistical downscaling does not need to directly use high-resolution dynamic models. For example, weather forecast services have often used linear regression or simple machine learning as postprocessing guidance to obtain detailed, or town-by-town, forecasts (cf. Glahn et al. 2009). Climate researchers have also used statistical downscaling to predict regional climate change with lower computational cost (cf. Gutiérrez et al. 2018).

On the other hand, as machine learning techniques have evolved, a new, promising method has emerged: a statistical downscaling algorithm, that is, single-image super resolution (SISR), which originally estimates high-resolution images from low-resolution counterparts using powerful deep neural networks (cf. Yang et al. 2019; Wang et al. 2020). Recently, SISR has been used for the climatological or daily estimation of precipitation (e.g., Weber et al. 2020; Baño-Medina et al. 2020). The advantage of SISR is the ability to construct high-resolution gridded values, not only values at specific points (cities or observatories). Therefore, if the SISR downscaling calculation is performed with a short time interval (hours), the calculation continuously provides gridded meteorological fields available for numerical weather prediction (NWP). That is a surrogate model capable of simulating mesoscale or microscale advection, diffusion, deposition, cloud physics, and land surface processes (cf. Reichstein et al. 2019).

However, to the best of our knowledge, the SISR algorithm has never been publicly applied to mesoscale surrogate modeling to simulate meteorological gridded fields. Only one exception is the conference presentation by Kern et al. (2020). Generally, low-resolution (or synoptic-scale) dynamic models can never adequately simulate high-resolution (or mesoscale) advection, diffusion, and deposition fields over complex terrain, as shown by Sekiyama et al (2015; 2017) and Sekiyama and Kajino (2020). Fundamentally, low-resolution models do not retain information on complex terrain below their resolution. Therefore, we will benefit greatly from downscaling surrogate modeling to compensate for the missing information on complex terrain and terrain-driven weather patterns. SISR downscaling will provide high-resolution gridded meteorological analyses, which can drive pollutant transport models. These low-cost downscaling models are especially beneficial for prompt responses to local environmental emergencies such as nuclear power plant accidents.

In this study, we show a preliminary experiment of such a downscaling surrogate model from the synoptic scale to the mesoscale. The SISR downscaling algorithm was applied to estimate 5-km gridded surface temperature fields from 22-km gridded surface temperature and wind fields. This was a simplified case of surrogate modeling for NWP in the planetary boundary layer. The estimation was performed at 6-hour intervals, not daily averaged or seasonal intervals, as in previous studies for climatology. The estimation domain covered central Japan, which contains mountains, highlands, basins, valleys, peninsulas, water bodies, and small islands. We demonstrate the downscaling performance of SISR, or deep convolutional neural networks (DCNNs), over complex terrain.

\begin{figure}[ht]
  \begin{center}
    \includegraphics[width=120mm]{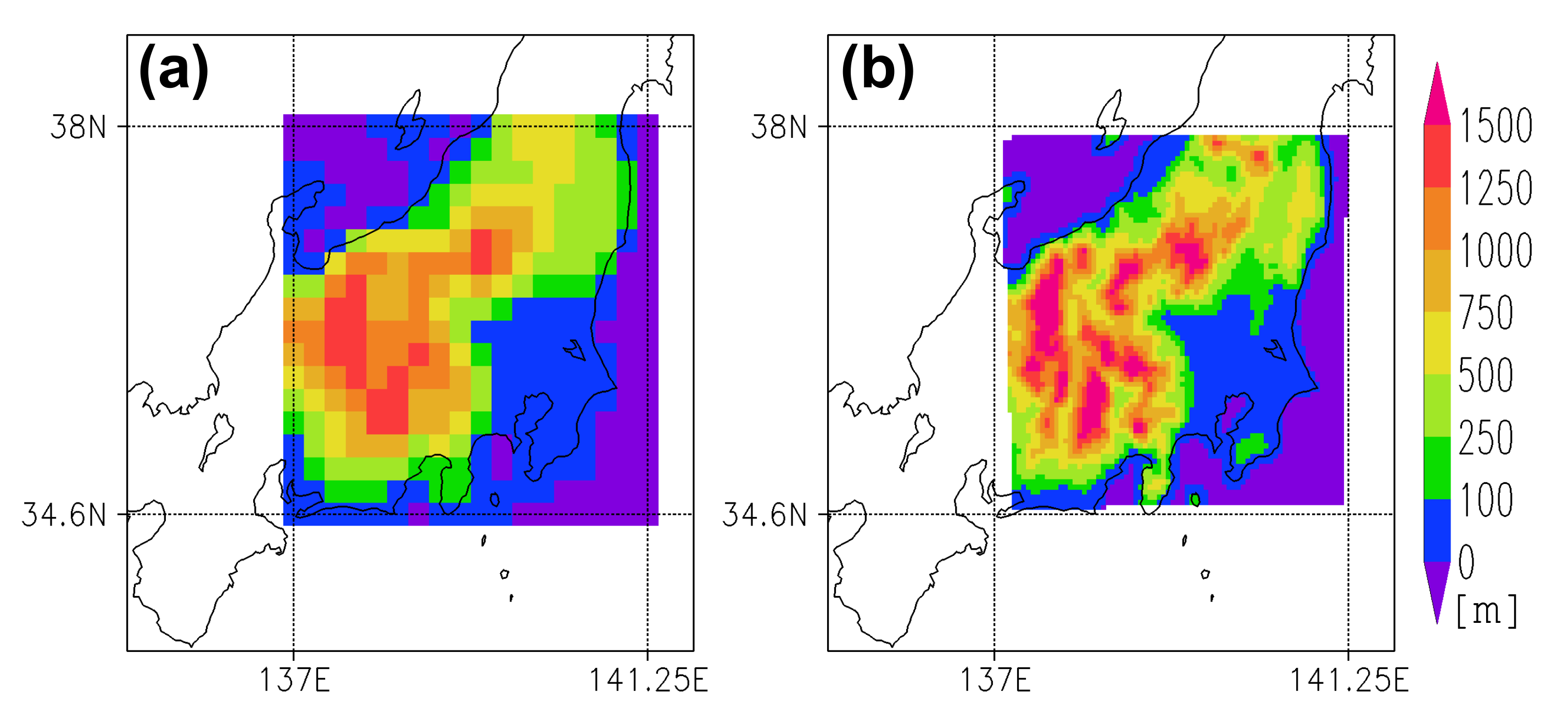}
  \end{center}
  \caption{Model domains and resolutions for (a) the 22-km gridded inputs and (b) the 5-km gridded outputs. The colors indicate the elevations represented in each analysis.}
\end{figure}

\begin{figure}[ht]
  \begin{center}
    \includegraphics[width=70mm]{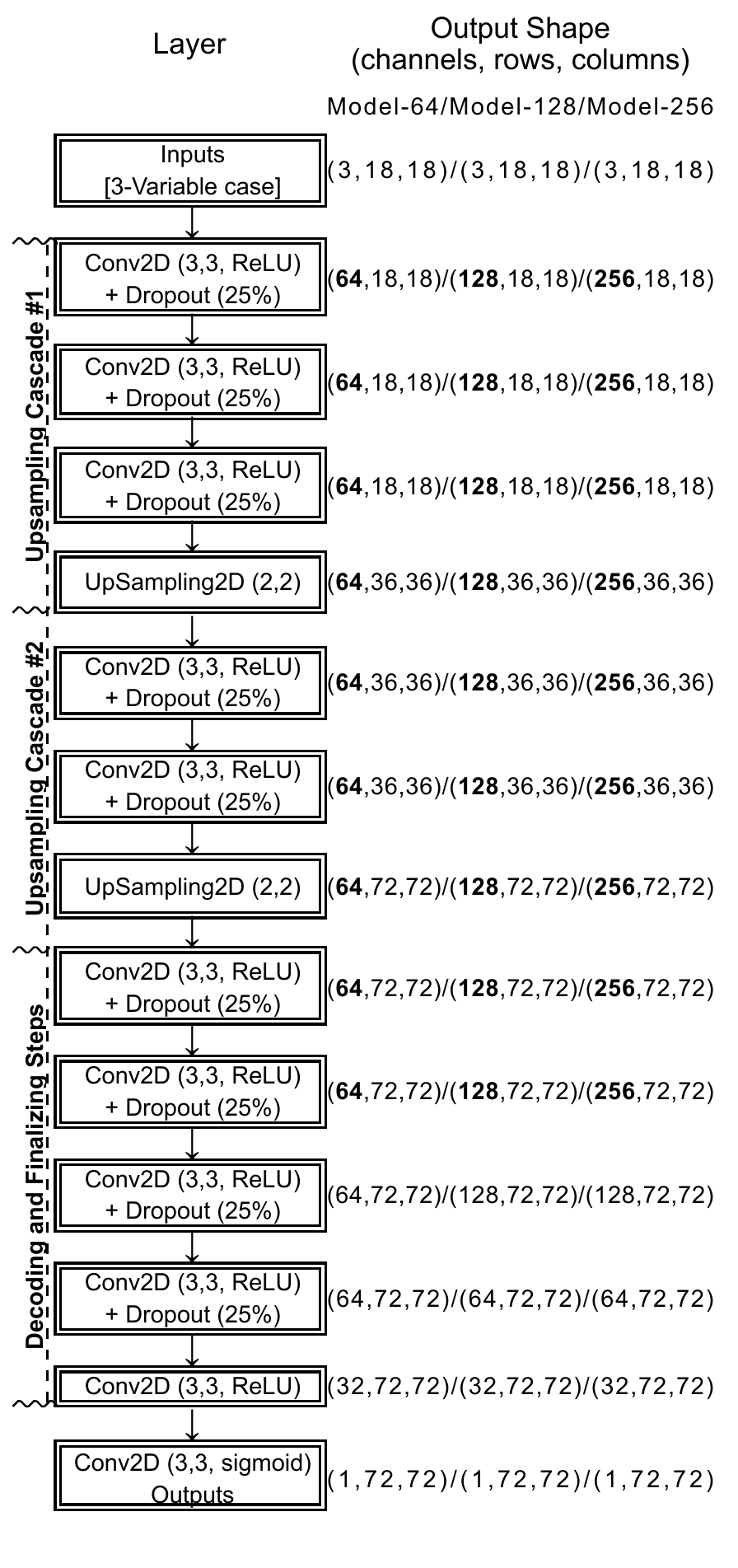}
  \end{center}
  \caption{Schematic diagram of the neural network for our SISR models.}
\end{figure}

\begin{figure}[ht]
  \begin{center}
    \includegraphics[width=60mm]{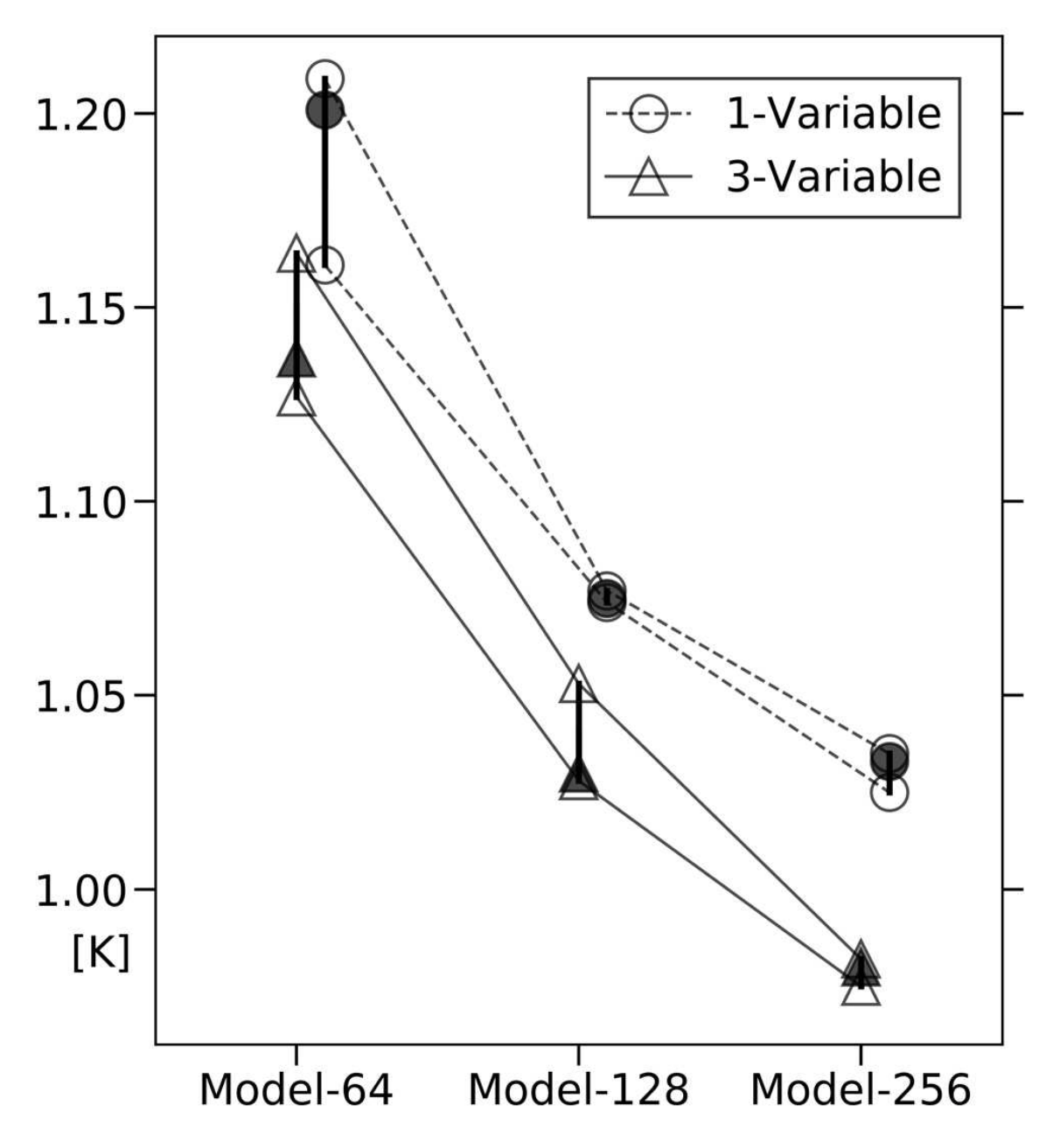}
  \end{center}
  \caption{Root-mean-square errors of the model outputs over the entire model domain. The open circles/triangles indicate the maximum and minimum scores of each model. The filled circles/triangles indicate the median scores of each model.}
\end{figure}

\section{Data and Methods}
Prior to performing SISR downscaling, we prepared input and desired output files of meteorological analyses. The input file consists of temperatures at a height of 2 m (hereafter, called T2), zonal winds at a height of 10 m and meridional winds at a height of 10 m (hereafter, called U10 and V10, respectively). The inputs are fixed-size 18 × 18 pixel images with a 0.25 × 0.20 degree (approximately 22 km) horizontal resolution covering longitudinally from 137.00 $^\circ$E to 141.25 $^\circ$E and latitudinally from 34.60 $^\circ$N to 38.00 $^\circ$N (Fig. 1a). The inputs are 6-hourly data extracted from the Japan Meteorological Agency (JMA) operational global analysis (cf. World Meteorological Organization 2015) without any interpolation. The desired output file consists of only temperature data (T2), which are constructed into 72 × 72 pixels with an exactly 5-km horizontal resolution in the Lambert coordinate system covering almost the same area as the inputs but slightly smaller (Fig. 1b). The desired outputs are 6-hourly data extracted from the JMA operational mesoscale analysis (cf. Honda et al. 2005) without any interpolation.

Both the inputs and desired outputs are separated into a 4-year training period (April 1, 2015 – March 31, 2019) and a 1-year test period (April 1, 2019 – March 31, 2020). All meteorological fields (i.e., T2, U10, and V10 for the inputs and T2 for the desired outputs) are snapshotted four times per day at 00:00, 06:00, 12:00, and 18:00 UTC according to the operational global analysis cycle. The local time (LT) is 9 hours ahead of UTC. The bias in the input T2 and desired output T2 data are removed by subtracting the 31-day running mean temperature of the input T2 averaged over the 18 × 18 pixels horizontally and for the same day and time from April 2015 to March 2020. The unbiased T2 values are normalized by setting the maximum value to +12 K and the minimum value to -12 K, dividing by 24 K, and adding 0.5. The final values of T2 range from 0 to 1, where the maximized or minimized pixels are 0.04 \% in total for the input and 0.5 \% in total for the desired output. The inputs U10 and V10 are normalized by using a logistic sigmoid function with a gain of 0.2 that makes the sigmoid curve nearly linear between -10 ms$^{-1}$ and +10 ms$^{-1}$.

SISR downscaling models were constructed by using a progressive upsampling framework (cf. Chapter 3 of Wang et al. 2020) of DCNNs and large-scale image recognition (cf. Simonyan and Zisserman 2015), as shown in Fig. 2. There are two upsampling cascades, each of which magnifies an input image by 2 × 2 times. Therefore, 18 × 18 pixel inputs become 72 × 72 pixel outputs obtained through the two-stage progressive upsampling procedure. The model source codes were written in TensorFlow 2.2.0 (Ababi et al. 2015) and Keras 2.3.1 (Chollet 2015) via Python. The application programming interface (API) parameters of TensorFlow and Keras were set to the default values if not otherwise specified in this section. All the convolutional layers, except for the output layer, had a kernel size of 3 × 3, a stride of 1, the same padding, and the rectified linear unit (ReLU) activation function. The output layer was constructed by using a convolutional layer with the same settings as the other convolutional layers, except the standard sigmoid activation function was. All the dropout layers had a dropout rate of 25 \%. The weights and biases in the DCNN were optimized by employing the mean of the absolute errors as the loss function and by using the Adam optimizer (Kingma and Ba 2014) with a learning rate of 0.001. The models were trained with a batch size of 128 and 100 epochs. These hyperparameters were tuned through trial and error without test data.

We prepared three SISR models by varying the number of filters (output channels) of the main convolutional layers, as shown in Fig. 2. One has 64 filters, one has 128 filters, and the other has 256 filters for each convolutional layer. Hereafter, we call them Model-64, Model-128, and Model-256, respectively. Then, we performed two types of experiments to evaluate the effect of wind fields on temperature fields. One experiment involves only temperature (T2) fields in the inputs, and the other involves not only temperature (T2) but also wind (U10 and V10) fields. Hereafter, we call them 1-Variable and 3-Variable, respectively. In addition, we performed two types of training methods to evaluate the effect of the time slot of the inputs/outputs. One method uses all-day time slots for simultaneous training, but the other uses each time slot (00:00, 06:00, 12:00, or 18:00 UTC) separately.

\begin{table}[ht]
	\caption{Statistics for the 1-year test period of the 22-km gridded inputs and 5-km gridded outputs using the 1-Variable or 3-Variable Model-256.}
	\begin{center}
	\begin{tabular}{cc|rrr}
		\ & & Input & \shortstack{1-Var.\\Output} & \shortstack{3-Var.\\Output} \\
		\hline
		\multirow{5}{*}{RMSE [K]}
		& Total & 1.89 & 1.02 & 0.98 \\
		& Water bodies & 1.46 & 0.94 & 0.90 \\
		& Lowlands & 1.57 & 1.01 & 1.01 \\
		& Midlands & 1.97 & 1.04 & 1.00 \\
		& Highlands & 2.69 & 1.15 & 1.04 \\
		\hline
		\multirow{5}{*}{Bias [K]}
		& Total & 0.29 & -0.09 & 0.12 \\
		& Water bodies & 0.08 & -0.18 & -0.03 \\
		& Lowlands & 0.73 & 0.09 & 0.30 \\
		& Midlands & -0.07 & -0.07 & 0.21 \\
		& Highlands & 1.36 & -0.15 & 0.00 \\
		\hline
		\multirow{5}{*}{Correlation}
		& Total & 0.73 & 0.87 & 0.88 \\
		& Water bodies & 0.83 & 0.94 & 0.94 \\
		& Lowlands & 0.73 & 0.81 & 0.82 \\
		& Midlands & 0.70 & 0.84 & 0.86 \\
		& Highlands & 0.64 & 0.85 & 0.88
	\end{tabular}
	\label{tab:1}
	\end{center}
\end{table}

\begin{table}[ht]
	\caption{RMSEs [K] for each time slot.}
	\begin{center}
	\begin{tabular}{cc|rrr}
		\ & & Input & \shortstack{1-Var.\\Output} & \shortstack{3-Var.\\Output} \\
		\hline
		\multirow{4}{*}{Water bodies}
		& 00:00 UTC & 1.38 & 0.99 & 0.91 \\
		& 06:00 UTC & 1.68 & 0.95 & 0.96 \\
		& 12:00 UTC & 1.28 & 0.88 & 0.84 \\
		& 18:00 UTC & 1.43 & 0.96 & 0.88 \\
		\hline
		\multirow{4}{*}{Lowlands}
		& 00:00 UTC & 1.36 & 1.02 & 1.03 \\
		& 06:00 UTC & 1.45 & 0.99 & 1.00 \\
		& 12:00 UTC & 1.77 & 0.99 & 0.96 \\
		& 18:00 UTC & 1.66 & 1.05 & 1.04 \\
		\hline
		\multirow{4}{*}{Midlands}
		& 00:00 UTC & 1.85 & 1.00 & 0.98 \\
		& 06:00 UTC & 1.91 & 0.99 & 0.98 \\
		& 12:00 UTC & 1.89 & 1.00 & 0.96 \\
		& 18:00 UTC & 2.20 & 1.17 & 1.06 \\
		\hline
		\multirow{4}{*}{Highlands}
		& 00:00 UTC & 2.53 & 1.11 & 1.03 \\
		& 06:00 UTC & 3.23 & 1.05 & 0.98 \\
		& 12:00 UTC & 2.47 & 1.13 & 1.01 \\
		& 18:00 UTC & 2.46 & 1.29 & 1.14
	\end{tabular}
	\label{tab:2}
	\end{center}
\end{table}

\begin{table}[ht]
	\caption{RMSEs [K] for each season.}
	\begin{center}
	\begin{tabular}{cc|rrr}
		\ & & Input & \shortstack{1-Var.\\Output} & \shortstack{3-Var.\\Output} \\
		\hline
		\multirow{4}{*}{Water bodies}
		& DJF & 1.59 & 0.98 & 0.98 \\
		& MAM & 1.38 & 0.80 & 0.81 \\
		& JJA & 1.37 & 0.99 & 0.90 \\
		& SON & 1.51 & 1.00 & 0.88 \\
		\hline
		\multirow{4}{*}{Lowlands}
		& DJF & 1.77 & 1.10 & 1.11 \\
		& MAM & 1.36 & 0.86 & 0.85 \\
		& JJA & 1.42 & 1.01 & 0.95 \\
		& SON & 1.68 & 1.06 & 1.11 \\
		\hline
		\multirow{4}{*}{Midlands}
		& DJF & 2.04 & 1.14 & 1.11 \\
		& MAM & 1.68 & 0.88 & 0.85 \\
		& JJA & 1.99 & 1.06 & 0.98 \\
		& SON & 2.13 & 1.09 & 1.03 \\
		\hline
		\multirow{4}{*}{Highlands}
		& DJF & 3.02 & 1.24 & 1.14 \\
		& MAM & 2.59 & 0.93 & 0.87 \\
		& JJA & 2.45 & 1.18 & 1.07 \\
		& SON & 2.69 & 1.22 & 1.07
	\end{tabular}
	\label{tab:3}
	\end{center}
\end{table}

\begin{table}[ht]
	\caption{RMSEs [K] for the all-day trained result and the time-slot trained result using 3-Variable Model-256.}
	\begin{center}
	\begin{tabular}{cc|cc}
		\ & & \shortstack{All-day\\trained} & \shortstack{Time-slot\\trained} \\
		\hline
		\multirow{4}{*}{Water bodies}
		& 00:00 UTC & 0.91 & 0.94 \\
		& 06:00 UTC & 0.96 & 0.97 \\
		& 12:00 UTC & 0.84 & 0.88 \\
		& 18:00 UTC & 0.88 & 0.91 \\
		\hline
		\multirow{4}{*}{Lowlands}
		& 00:00 UTC & 1.03 & 1.03 \\
		& 06:00 UTC & 1.00 & 1.00 \\
		& 12:00 UTC & 0.96 & 1.02 \\
		& 18:00 UTC & 1.04 & 1.03 \\
		\hline
		\multirow{4}{*}{Midlands}
		& 00:00 UTC & 0.98 & 1.01 \\
		& 06:00 UTC & 0.98 & 0.99 \\
		& 12:00 UTC & 0.96 & 1.05 \\
		& 18:00 UTC & 1.06 & 1.09 \\
		\hline
		\multirow{4}{*}{Highlands}
		& 00:00 UTC & 1.03 & 1.04 \\
		& 06:00 UTC & 0.98 & 1.01 \\
		& 12:00 UTC & 1.01 & 1.14 \\
		& 18:00 UTC & 1.14 & 1.20
	\end{tabular}
	\label{tab:4}
	\end{center}
\end{table}

\section{Results}
In this study, the root-mean-square errors (RMSEs) and biases were evaluated by calculating a snapshot RMSE or bias spatially over the 72 × 72 pixel outputs at each time/day and then averaging the snapshots over the test period. In contrast, the correlation was evaluated by calculating the Pearson correlation coefficient temporally for the test period for each pixel and then averaging the correlation coefficients over the 72 × 72 pixel outputs. The statistics of the low-resolution inputs were estimated by nearest-neighbor interpolation on the high-resolution output coordinate system. The RMSEs were calculated not only for the entire model domain but also for each terrain, i.e., lowlands, midlands, highlands, and water bodies. Water bodies were identified by a water/land ratio greater than 0.5 in the 5-km gridded model. We defined lowlands as areas at less than 100 m elevation, highlands as areas at more than 1000 m, and midlands as areas with elevations between 100 m and 1000 m in the 5-km gridded model (cf. Fig. 1b). As a result, we had 1633 pixels of water bodies, 773 pixels of lowlands, 2052 pixels of midlands, and 726 pixels of highlands.

We performed 4-year model training and 1-year test evaluation three times for each model (1- or 3-Variable and Model-64, 128, or 256) with various random seeds. The test results are shown in Fig. 3, indicating that there was a significant difference between the model outputs. Note that the RMSE of the model inputs was 1.89 K for the 1-year test period (cf. Table 1). The difference between the model outputs (approximately 0.2 K) was smaller than the improvement from the model inputs (approximately 1 K). Nevertheless, the best performance was clearly presented by Model-256 with a small spread of the three ensemble members. Hereafter, we illustrate only the best member of 1- or 3-Variable Model-256.

Details of the statistics for the 1-year test period are tabulated in Table 1. The RMSEs indicated that (1) the scores of the input data substantially deteriorated as the elevation increased, (2) but the scores of the output data were almost equalized, (3) and consequently, the score was improved by 1.7 K for the highlands, although it improved by only 0.5 K for the water bodies. The bias of the input data was relatively large, especially in the highlands (1.4 K), while the bias in the output data was relatively small (less than ±0.3 K). The correlation for the input data slightly deteriorated as the elevation increased. In contrast, for the output data, the correlation slightly improved as the elevation increased. The correlation coefficient was improved from 0.6–0.7 to 0.8–0.9 on land by the model.

We divided the RMSEs into four time slots (Table 2). Generally, the temporal variability in the RMSE was relatively small except for the input data for the highlands. In the highlands, the worst performance occurred in the afternoon local time (06:00 UTC = 15:00 LT) for the input data but late at night local time (18:00 UTC = 03:00 LT) for the output data. Conversely, the best performance occurred late at night for the input data but in the afternoon for the output data. Comparing the 1-Variable and 3-Variable model outputs, improvement was negligible for the water bodies, lowlands, and midlands. However, the 3-Variable model was clearly superior to the 1-Variable model for the highlands, especially at night local time (12:00 and 18:00 UTC = 21:00 and 03:00 LT, respectively).

We divided the RMSEs into four seasons (Table 3): December, January, and February (DFJ; winter), March, April, and May (MAM; spring), June, July, and August (JJA; summer), and September, October, and November (SON; autumn). Improvement from the 1-Variable to the 3-Variable model was negligible except for the highlands, although the 3-Variable model was slightly superior for the water bodies in summer (JJA) and autumn (SON). Additionally, the effect of time-slot training is shown in Table 4. Unfortunately, no improvement was achieved through time-slot training. Instead, the scores became slightly worse in all areas.

Finally, examples of the diurnal temperature (T2) distributions are demonstrated in Fig. 4. These snapshots (at 00:00, 06:00, 12:00, and 18:00 UTC on December 29, 2019) were chosen by their moderate RMSEs (close to the averages) and the distinct diurnal change in temperature on land. Note that the inputs cannot represent borders between land and water bodies when the contrast was faint. While the inputs could represent principal mountains in central Japan, any valleys and basins in the mountains are not represented at all. Consequently, we can hardly distinguish the shape of central Japan in the input maps. In contrast, the model outputs clearly drew the landscape of Japan as the desired outputs did. Although any small islands, peninsulas, or bays do not exist in the input maps, they emerge in the model output maps (Tokyo Bay, for instance). Since detailed mountain ridges are represented in the model outputs, we can identify small valleys and basins that do not exist in the inputs.

\section{Discussion}
Concerning the design of our SISR models, the effect of layer-size augmentation on model performance was significantly positive, as shown in Fig. 3. However, the effect seemed very small and almost saturated near 256 filters per layer. On the other hand, our SISR models did not appropriately work when reducing the depth of neural networks to or upsampling cascades. Previous studies such as those by Yang et al. (2019), Weber et al. (2020), and Wang et al. (2020) recommend using neural networks as deep as ours or deeper for large-scale image recognition. Fortunately, our model results seemed adequate enough to convert synoptic-scale analyses to mesoscale analyses. In Fig. 4, the model outputs were not calculated from the training data; they were calculated from the test data. However, the locations of coastlines and mountain ridges were identified consistently with the desired outputs. When the sea surface temperature (SST) was much higher than the land surface temperature (LST), the coastlines were easily identified, as seen on the northwest coast of central Japan in Fig. 4c or 4d. However, even if the SST was very close to the LST, the coastlines seemed precisely reproduced, for example, on the south coast of central Japan, as shown in Fig. 4a, 4b, 4c, or 4d. The model performance, i.e., RMSE $\approx$ 1 K, bias $\approx$ 0 K, and correlation $\approx$ 0.9, would be worth using for operational weather forecast guidance (cf. Glahn et al. 2009) or real-time air pollution alerts. Temperature is one of the most important factors of atmospheric chemical reactions.

As shown in Table 1, the RMSEs and biases of the inputs significantly deteriorated as the elevation increased. This finding is reasonable because temperature strongly depends on elevation, and there is a large discrepancy at high elevations between low resolutions and high resolutions over complex terrain. For example, Mt. Fuji does not exist in the 22-km gridded inputs. In contrast, the RMSEs of the outputs rarely depended on the elevation. This finding indicates that the SISR model was evenly well trained. Note that these scores were also improved in water bodies, which might be caused by (1) the improvement in the reproducibility of the coastlines by high-resolution pixels and (2) the difference between the nature of the input analysis and the nature of the desired output analysis. The difference between the 1-Variable model and the 3-Variable model seemed small in these scores. The wind dependency will be much weaker than the elevation dependency for temperature, while the temperature dependency for wind is large in general. In addition, since remarkable wind-dependent phenomena, such as foehn winds, are not common, they might not be well trained or not be detectable when averaged.

Time dependency was relatively small for the water bodies, lowlands, and midlands, as shown in Table 2. The inputs at the highlands deteriorated in the afternoon (06:00 UTC = 15:00 LT) and improved late at night (18:00 UTC = 03:00 LT). In other words, the elevation dependency of temperature was probably larger in the afternoon. This phenomenon might be caused by solar radiation heating maximization at noon. In contrast, the outputs for the highlands (and probably for the midlands) slightly deteriorated late at night (18:00 UTC = 03:00 LT), probably because the surface air often cools at night and gathers in valleys or basins. This temperature gradient is in the direction opposite to the regular gradient along the elevation. Nevertheless, relatively large improvements by the 3-Variable model were observed late at night in the midlands/highlands. This finding perhaps indicates that the night cooling tends to be influenced by surface winds. In addition, radiation cooling at night strongly depends on the weather. Therefore, augmentation of the input variables (e.g., cloud coverage or cloud height) will be needed in future studies.

The seasonal change in performance was small for both the inputs and outputs (Table 3). In addition, only the inputs presented a poor performance in winter (DJF) for the highlands. This phenomenon might be caused by snowfall that is never well reproduced by low-resolution dynamic models. The 3-Variable slightly improved the downscaling performance in summer (JJA) and autumn (SON) for the water bodies (Table 3), which is probably due to typhoons and monsoons coming to Japan mainly in these seasons. In Table 4, time-slot training slightly worsened the downscaling performance, which was probably caused by a shortage of training samples. Since the number of time-slot samples was one-fourth of the total, we might need 16-year-long inputs and desired outputs for fairness. However, it is not easy to obtain mesoscale analysis data with the 16-year-long coherent quality and bias of a dynamic model and a data assimilation system.

Note that the desired outputs are not observations but the computational results of a mesoscale dynamic model. Hence, our SISR models imitate not the real world but the mesoscale dynamic model. Nevertheless, the imitation of the dynamic model is our very objective in this study because gridded analyses are definitely needed to drive nested models, such as pollutant transport models. Furthermore, such nested models require meteorological variables and vertical profiles other than the surface temperature. Therefore, our SISR models will be augmented to include vertical multilayers (e.g., not only the surface but also 900, 850, and 700 hPa) and multiple variables (e.g., wind, pressure, precipitation, and cloud coverage). In addition, recurrent neural networks (cf. Lipton et al. 2015) might be beneficial to meteorological downscaling because weather changes are temporally sequential.

\begin{figure}[ht]
  \begin{center}
    \includegraphics[width=140mm]{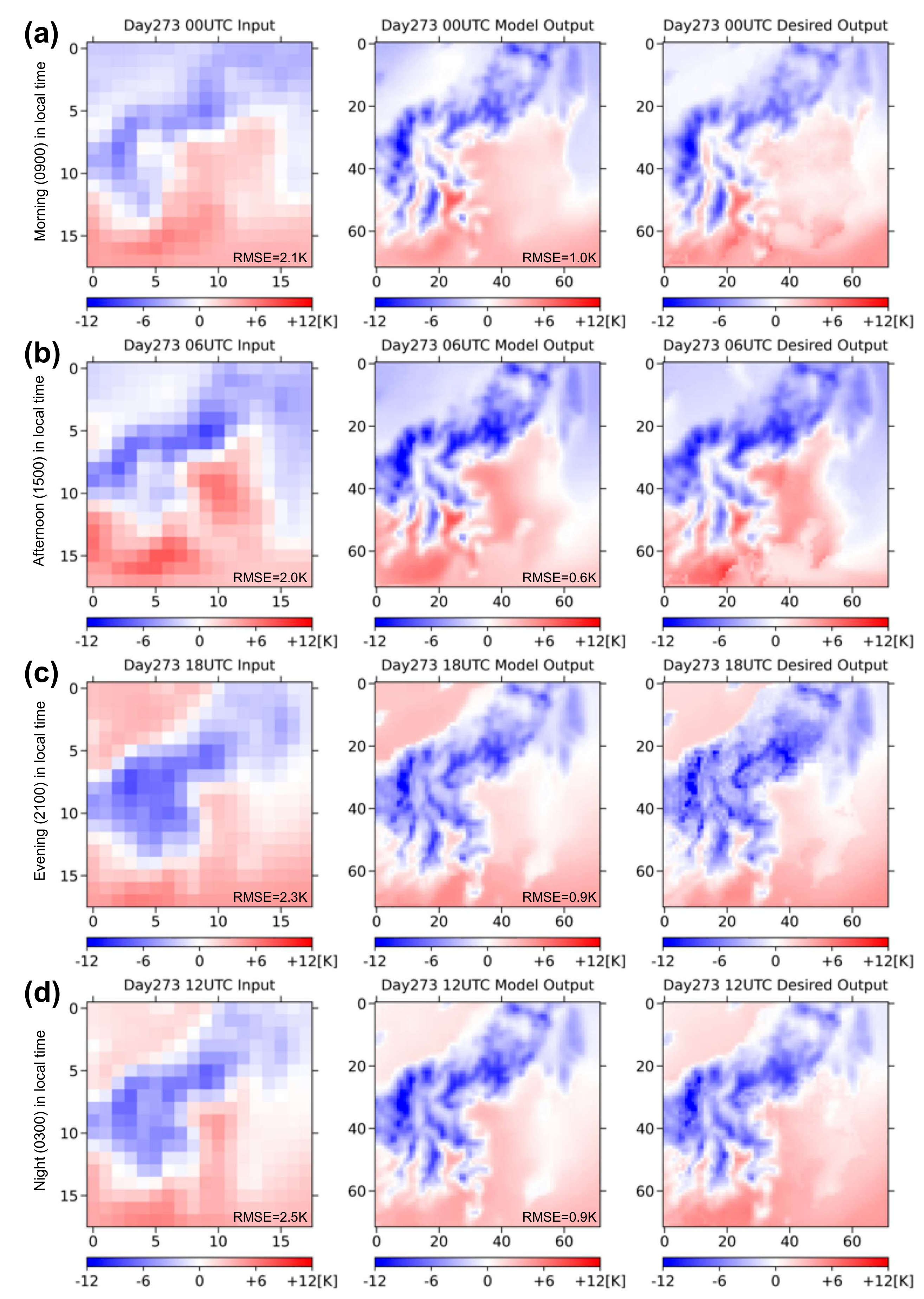}
  \end{center}
  \caption{Unbiased surface temperature (T2) snapshots from December 29, 2019, at (a) 00:00 UTC or 09:00 local time, (b) 06:00 UTC or 15:00 local time, (c) 12:00 UTC or 21:00 local time, and (d) 18:00 UTC or 03:00 local time. The snapshots are taken from the inputs of the test data (left), the model outputs using 3-Variable Model-256 (middle), and the desired outputs of the test data (right).}
\end{figure}

\section{Conclusions}
We have described an experiment for a surrogate model to downscale surface temperature fields every 6 hours from the synoptic scale to the mesoscale. We used deep convolutional neural networks (DCNNs) known as single-image super resolution (SISR). We found that the SISR technique was capable of downscaling the temperature correctly enough for surrogate modeling. While we downscaled only the surface temperature in this study, future experiments will be performed with other prognostic variables and vertical profiles. Generally, DCNNs do not require much computational power once training is finished. Therefore, SISR surrogate models can be used for short time intervals or in emergency situations. They will enable high-resolution weather forecasts or environmental emergency responses (EERs). Thus, we like to implement pollutant transport models (PTMs) by inputting the meteorological variables downscaled by the surrogate models. Moreover, some pioneering studies have developed DCNN models to directly predict plume advection and diffusion (de Bézenac et al. 2017; Tompson et al. 2017). This implies that the surrogate models might have the ability to directly simulate air pollution as high-resolution PTMs.

\section*{Acknowledgments}
The global and mesoscale meteorological analyses used in this study were provided operationally by the Japan Meteorological Agency via the Japan Meteorological Business Support Center (2020, updated daily, unpublished), which are free for research purposes. This work was supported by the Environment Research and Technology Development Fund (1-1802 and 5-2001) from the Environmental Restoration and Conservation Agency and by JSPS KAKENHI Grant Number JP17K00533 and JP19H01155.

\section*{References}
Abadi, M., A. Agarwal, P. Barham, E. Brevdo, Z. Chen, C. Citro, G. S. Corrado, A. Davis, J. Dean, M. Devin, S. Ghemawat, I. Goodfellow, A. Harp, G. Irving, M. Isard, R. Jozefowicz, Y. Jia, L. Kaiser, M. Kudlur, J. Levenberg, D. Mané, M. Schuster, R. Monga, S. Moore, D. Murray, C. Olah, J. Shlens, B. Steiner, I. Sutskever, K. Talwar, P. Tucker, V. Vanhoucke, V. Vasudevan, F. Viégas, O. Vinyals, P. Warden, M. Wattenberg, M. Wicke, Y. Yu, and X. Zheng, 2015: TensorFlow: Large-scale machine learning on heterogeneous systems, https://www.tensorflow.org.
\vspace{\baselineskip}

\noindent
Baño-Medina, J., R. Manzanas, and J. M. Gutiérrez, 2020: Configuration and intercomparison of deep learning neural models for statistical downscaling, \textit{Geosci. Model Dev.}, 13, 2109–2124, doi:10.5194/gmd-13-2109-2020.
\vspace{\baselineskip}

\noindent
de Bézenac, E., A. Pajot, and P. Gallinari, 2017: Deep learning for physical processes: incorporating prior scientific knowledge, arXiv:1711.07970.
\vspace{\baselineskip}

\noindent
Chollet, F., 2015: Keras, https://github.com/fchollet/keras.
\vspace{\baselineskip}

\noindent
Glahn, B., K. Gilbert, R. Cosgrove, D. P. Ruth, and K. Sheets, 2009: The Gridding of MOS, \textit{Wea. Forecasting}, 24, 520–529, doi:10.1175/2008WAF2007080.1.
\vspace{\baselineskip}

\noindent
Gutiérrez, J. M., D. Maraun, M. Widmann, R. Huth, E. Hertig, R. Benestad, O. Roessler, J. Wibig, R. Wilcke, S. Kotlarski, D. San Martín, S. Herrera, J. Bedia, A. Casanueva, R. Manzanas, M. Iturbide, M. Vrac, M. Dubrovsky, J. Ribalaygua, J. Pórtoles, O. Räty, J. Räisänen, B. Hingray, D. Raynaud, M. J. Casado, P. Ramos, T. Zerenner, M. Turco, T. Bosshard, P. Štěpánek, J. Bartholy, R. Pongracz, D. E. Keller, A. M. Fischer, R. M. Cardoso, P. M. M. Soares, B. Czernecki, and C. Pagé, 2018: An intercomparison of a large ensemble of statistical downscaling methods over Europe: Results from the VALUE perfect predictor cross‐validation experiment, \textit{Int. J. Climatol.}, 39, 3750–3785, doi:10.1002/joc.5462.
\vspace{\baselineskip}

\noindent
Honda, Y., M. Nishijima, K. Koizumi, Y. Ohta, K. Tamiya, T. Kawabata, and T. Tsuyuki, 2005: A pre-operational variational data assimilation system for a non-hydrostatic model at the Japan Meteorological Agency: Formulation and preliminary results, \textit{Quart. J. Roy. Meteor. Soc.}, 131, 3465–3475, doi:10.1256/qj.05.132.
\vspace{\baselineskip}

\noindent
Kern, M. K. Höhlein, T. Hewson, and R. Westermann, 2020: Towards Operational Downscaling of Low Resolution Wind Fields using Neural Networks, \textit{EGU General Assembly 2020}, EGU2020-5447, doi:10.5194/egusphere-egu2020-5447.
\vspace{\baselineskip}

\noindent
Kingma, D. P. and Ba, J., 2014: Adam: A Method for Stochastic Optimization, arXiv:1412.6980.
\vspace{\baselineskip}

\noindent
Lipton, Z. C., J. Berkowitz, and C. Elkan, 2015: A Critical Review of Recurrent Neural Networks for Sequence Learning, arXiv:1506.00019.
\vspace{\baselineskip}

\noindent
Reichstein, M., G. Camps-Valls, B. Stevens, M. Jung, J. Denzler, N. Carvalhais, and Prabhat, 2019: Deep learning and process understanding for data-driven Earth system science, \textit{Science}, 566, 195–204, doi:10.1038/s41586-019-0912-1.
\vspace{\baselineskip}

\noindent
Sekiyama, T. T., M. Kunii, K. Kajino, and T. Shimbori, 2015: Horizontal Resolution Dependence of Atmospheric Simulations of the Fukushima Nuclear Accident Using 15-km, 3-km, and 500-m Grid Models, \textit{J. Meteor. Soc. Japan}, 93, 49–64, doi:10.2151/jmsj.2015-002.
\vspace{\baselineskip}

\noindent
Sekiyama, T. T., M. Kunii, and K. Kajino, 2017: The Impact of Surface Wind Data Assimilation on the Predictability of Near-Surface Plume Advection in the Case of the Fukushima Nuclear Accident, \textit{J. Meteor. Soc. Japan}, 95, 447–454, doi:10.2151/jmsj.2017-025.
\vspace{\baselineskip}

\noindent
Sekiyama, T. T. and M. Kajino, 2020: Reproducibility of Surface Wind and Tracer Transport Simulations over Complex Terrain Using 5-, 3-, and 1-km-Grid Models, \textit{J. Appl. Meteor. Climatol.}, 59, 937–952, doi:10.1175/JAMC-D-19-0241.1.
\vspace{\baselineskip}

\noindent
Simonyan, K. and A. Zisserman, 2015: Very Deep Convolutional Networks for Large-Scale Image Recognition, arXiv:1409.1556v6.
\vspace{\baselineskip}

\noindent
Tompson, J., K. Schlachter, P. Sprechmann, and K. Perlin, 2017: Accelerating Eulerian fluid simulation with convolutional networks, arXiv:1607.03597v6.
\vspace{\baselineskip}

\noindent
Wang, Z., J. Chen, and S. C. H. Hoi, 2020: Deep Learning for Image Super-resolution: A Survey, arXiv:1902.06068v2.
\vspace{\baselineskip}

\noindent
Weber, T., A. Corotan, B. Hutchinson, B. Kravitz, and R. Link, 2020: Technical note: Deep learning for creating surrogate models of precipitation in Earth system models, \textit{Atmos. Chem. Phys.}, 20, 2303–2317, doi:10.5194/acp-20-2303-2020.
\vspace{\baselineskip}

\noindent
World Meteorological Organization, 2015: Japan 2015, WMO Technical Progress Report on the Global Data Processing and Forecasting System and Numerical Weather Prediction Research 2015, World Meteorological Organization, 56 p.
\vspace{\baselineskip}

\noindent
Yang, W., X. Zhang, Y. Tian, W. Wang, J. H. Xue, and Q. Liao, 2019: Deep Learning for Single Image Super-Resolution: A Brief Review, arXiv:1808.03344v3.

\end{document}